\documentclass[aps,prd,nofootinbib,twocolumn,superscriptaddress,preprintnumbers,balancelastpage,longbibliography]{revtex4-1}
\usepackage{style}

\usepackage{graphicx}

\DeclareRobustCommand{\benchmarkstar}{%
  \raisebox{-1.5ex}[0pt][0pt]{%
    \includegraphics[height=4.5ex]{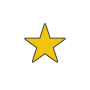}%
  }%
}
\DeclareRobustCommand{\benchmarkplus}{%
  \raisebox{-1.5ex}[0pt][0pt]{%
    \includegraphics[height=4.5ex]{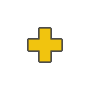}%
  }%
}

\begin{document}
\title{Cosmic-Ray Signals from Dark Matter Annihilation into Scalars with Baryon Number}

\author{Mattia Di Mauro\orcidlink{0000-0003-2759-5625}}
    \email{dimauro.mattia@gmail.com}
\affiliation{
    Istituto Nazionale di Fisica Nucleare, via P. Giuria, 1, 10125 Torino, Italy
}
\author{Caleb Gemmell\orcidlink{0000-0002-6505-8559}}
    \email{cgemmell2@wisc.edu}
\affiliation{
    Department of Physics, Wisconsin IceCube Particle Astrophysics Center, University of Wisconsin, Madison, Wisconsin 53706, USA
}
\author{Dan Hooper\orcidlink{0000-0001-8837-4127}}
    \email{dwhooper@wisc.edu}
\affiliation{
    Department of Physics, Wisconsin IceCube Particle Astrophysics Center, University of Wisconsin, Madison, Wisconsin 53706, USA
}
\author{Fabrizio Vassallo\orcidlink{0000-0002-0541-5606}}
    \email{fevassallo@wisc.edu}
\affiliation{
    Department of Physics, Wisconsin IceCube Particle Astrophysics Center, University of Wisconsin, Madison, Wisconsin 53706, USA
}

\begin{abstract}
Cosmic-ray antideuterons $\overline{\mathrm D}$ and antihelium-3 nuclei ${}^3\overline{\mathrm{He}}$ are promising probes of dark matter (DM). In light of potential detections of heavy antinuclei at AMS-02, we investigate a recently proposed scenario in which DM annihilates into color-singlet scalars carrying baryon and lepton number, which has been shown to enhance antideuteron and antihelium-3 fluxes relative to antiprotons. Accounting for Galactic propagation and detector sensitivities, we calculate the number of events predicted by our model for parameters that are compatible with the seven-year AMS-02 antiproton data. Broad regions of the DM-mediator mass plane result in observable antideuteron fluxes at AMS-02 and GAPS. While detecting ${}^3\overline{\mathrm{He}}$ remains more challenging, potentially detectable fluxes can arise for light mediators and moderately light DM, and in general are enhanced relative to the standard WIMP predictions. Furthermore, we find antihelium-to-antideuteron event ratios of $\mathcal{O}(10^{-1})$, closer to those suggested by the reported AMS-02 candidate events than conventional WIMP predictions.

\end{abstract}

\maketitle

%%%%%%%%%%%%%%%%%%%%%%%%%%%%%%%%%%%%%%%%%%%%%%%%
\section{Introduction}
\label{sec:intro}
%%%%%%%%%%%%%%%%%%%%%%%%%%%%%%%%%%%%%%%%%%%%%%%%

Understanding the particle description of dark matter (DM) has been an ongoing focus in the field of astroparticle physics and has motivated many studies of the potential indirect signatures of DM annihilation. DM annihilation generally produces equal amounts of matter and antimatter, whereas astrophysical mechanisms typically produce more matter than antimatter. Searches for antimatter in the cosmic-ray spectra are therefore a promising probe of DM. Importantly, astrophysical backgrounds of heavy antinuclei are small and suppressed at kinetic energies per nucleon below a few $\mathrm{GeV/n}$~\cite{Chardonnet:1997dv,Donato:1999gy,Cirelli:2014qia,Carlson:2014ssa,Stefanuto:2026wxy} due to kinematics, providing a particularly clean channel for DM detection.

No cosmic antideuteron or antihelium flux has yet been conclusively established. The strongest published antideuteron limit was obtained by BESS-Polar II,
\begin{equation}
\Phi_{\overline{\rm D}}
<
6.7\times10^{-5}\,
\mathrm{(m^2\,s\,sr\,GeV/n})^{-1},
\end{equation}
over an interval in their kinetic energy per nucleon of $K \in[0.163,1.100]\,\mathrm{GeV/n}$ \cite{BESS:2024yma}. An improvement in sensitivity of at least an order of magnitude is expected from AMS-02 and from the balloon-borne GAPS experiment, which recently completed its
first Antarctic science flight \cite{2008ICRC....4..765C,Aramaki:2015laa,Osteria:2020lxn,
vonDoetinchem:2020vbj,gaps_Hailey_cern}. These searches are particularly timely in light of the preliminary reports of several antideuteron detections and $\sim10$ antihelium events in the AMS-02 data \cite{Tingcern2016,Miapp2022DbarHebar,Miapp2022Dbar}.

If these events are confirmed, their relative rates would be difficult to explain with conventional DM models. Antinuclei form through the coalescence of antinucleons that are sufficiently close in both momentum and position space. Their formation probability consequently decreases rapidly with the mass number of the nucleus. For standard hadronic annihilation channels, event-by-event simulations typically yield
\begin{equation}
N_{\overline{p}} : N_\mathrm{\overline{D}} : N_{^3\mathrm{\overline{He}}}
\sim
1:10^{-4}:10^{-8},
\end{equation}
with some dependence on the annihilation channel, center-of-mass energy, and coalescence prescription~\cite{DiMauro:2024kml,DiMauro:2025vxp,
DiMauro:2026oto}. Increasing the annihilation cross section cannot generally compensate for this hierarchy because the much larger accompanying antiproton component is already tightly constrained by the precise measurements of AMS-02~ \cite{AMS:2021nhj,Korsmeier:2017xzj,Cuoco:2019kuu,
Calore:2022stf,Cholis:2019ejx}. An observable antihelium flux therefore requires a mechanism that enhances antinucleus formation relative to antiproton production, rather than simply increasing the total hadronic yield.

Several scenarios have been investigated in response to this tension. Within otherwise conventional WIMP models motivated by possible features in the antiproton spectrum, antideuteron fluxes can approach the projected sensitivities of AMS-02 and GAPS, while the corresponding antihelium flux generally remains below the sensitivity of
current experiments \cite{Cholis:2019ejx,Cholis:2020wum,
DeLaTorreLuque:2023antinuclei}. This has motivated more exotic models, both of DM \cite{Coogan:2017pwt,Heeck:2019ego,Winkler:2022zdu,Fedderke:2024hfy,DiMauro:2026suep} and other astrophysics \cite{Poulin:2018wzu,Bykov:2023nnr,Korwar:2024ofe}.

A novel scenario was recently proposed in Ref.~\cite{Gemmell:2026baryon}, highlighting that antinuclei fluxes can be enhanced in the decays of particles that carry baryon and lepton number. Such states can arise in extensions of the Standard Model with non-trivial baryon-number symmetries or in grand-unified constructions \cite{Kaplan:2009ag,Shelton:2010ta,Davoudiasl:2010am,FileviezPerez:2010gw,Davighi:2022qgb}. Thus, if these particles are produced through DM annihilation, they could potentially enhance the resulting antinuclei fluxes.

This scenario has two properties that favor antinucleus formation. First, each scalar decay injects several units of baryon or antibaryon number. Second, the energy is distributed among a larger number of hadronic systems, producing softer antinucleons with smaller characteristic phase-space separations. Ref.~\cite{Gemmell:2026baryon} found that these effects can enhance the antideuteron and antihelium-3 multiplicities by factors of approximately $10$--$100$ relative to conventional $\chi\chi\rightarrow b\bar b$ WIMP annihilation. Importantly, the ratios of antinuclei to antinucleons can also be enhanced, potentially allowing a large antinucleus signal without violating antiproton constraints from AMS-02.

In this work, we calculate the observable cosmic-ray signals of the baryon- and lepton-number-carrying scalar model of Ref.~\cite{Gemmell:2026baryon}. We generate the antiproton, antideuteron, and antihelium-3 source spectra for different DM and scalar masses using event-by-event hadronization and a coalescence model calibrated to collider data \cite{Bierlich:2022pfr,DiMauro:2024kml,DiMauro:2026oto,DiMauro:2026oto}. We then propagate these spectra to Earth using the semi-analytical framework adopted in Ref.~\cite{DiMauro:2026suep}. The seven-year AMS-02 antiproton dataset is used to constrain the annihilation cross section, and the corresponding antideuteron and antihelium-3 fluxes are compared with current limits and with the projected sensitivities of AMS-02 and GAPS. This allows us to establish whether the enhancement found at the production level leads to detectable antinuclei fluxes while remaining compatible with antiproton observations.

The remainder of this paper is organized as follows. In Sec.~\ref{sec:model_framework}, we summarize the DM model and its implementation to generate antinucleon spectra. Section~\ref{sec:coalescence} describes the coalescence model and the calculation of antinuclei spectra. The Galactic propagation setup is presented in Sec.~\ref{sec:propagation}. In Sec.~\ref{sec:results}, we derive the antiproton constraints, discuss antinucleus detection prospects at AMS-02 and GAPS, and assess a possible explanation of the Galactic Center excess. We summarize our conclusions in Sec.~\ref{sec:conclusion}.

%%%%%%%%%%%%%%%%%%%%%%%%%%%%%%%%%%%%%%%%%%%%%%%%
\section{Dark matter annihilation into scalars with baryon number}
\label{sec:model}
%%%%%%%%%%%%%%%%%%%%%%%%%%%%%%%%%%%%%%%%%%%%%%%%

\subsection{Dark Matter Model}
\label{sec:model_framework}

We consider the class of models developed in Ref.~\cite{Gemmell:2026baryon}, in which DM annihilates into color-singlet scalars carrying baryon and lepton number. Such states are motivated by extensions of the Standard Model in which baryon number is shared between the visible and dark sectors \cite{Kaplan:2009ag, Davoudiasl:2010am}, and by theories with spontaneously broken $U(1)_B$ and/or $U(1)_L$ symmetries \cite{Armbruster:2025jqs, FileviezPerez:2010gw}. Further motivation comes from unified theories that embed baryon–lepton symmetries in larger gauge groups \cite{Pati:1974yy, Davighi:2022qgb}. The $B=L=3$ assignment is particularly motivated by the electroweak structure of the Standard Model, which preserves $\mathbb{Z}_3$ subgroups of baryon and lepton number \cite{Koren:2022bam}. We focus on this assignment here and discuss the $B=L=1$ case in Appendix~\ref{app:other_scalars}. Recently, $U(1)_B$ extensions of the Standard Model have been invoked in endothermic and exothermic DM models to explain the high-energy LZ recoil event~\cite{DiMauro:2026syh,Gemmell:2026yaw}. In the present analysis, we do not specify a complete ultraviolet realization of the model, instead adopting a low-energy phenomenological description relevant for cosmic-ray production.

The self-conjugate scalar DM particle, $\chi$, which we assume to be stable on cosmological timescales, can annihilate to produce a pair of complex scalars,
\begin{equation}
\chi\chi\rightarrow
\phi_{B,L}\,\overline{\phi}_{B,L},
\label{eq:dm_phi_annihilation}
\end{equation}
where $m_{\rm DM} > m_\phi$ and $\phi_{B,L}$ has baryon and lepton numbers $(B,L)$, but is neutral under all SM gauge groups. The decay of the scalars proceeds as
\begin{align}
\phi_{3,3}
\rightarrow
3(udd\nu_e),
\qquad
\overline{\phi}_{3,3}
\rightarrow
3(\overline{udd\nu_e}).
\label{eq:phi33_decay}
\end{align}
While these specific decay products are chosen as a benchmark, other combinations of light quarks and leptons lead to qualitatively similar antinuclei yields \cite{Gemmell:2026baryon}. In particular, decay modes with a larger number of up quarks can moderately increase the ${}^3\overline{\mathrm{He}}$ yield at small $m_\phi$, but this difference decreases as the $\phi_{3,3}$ mass increases. This process corresponds schematically to the dimension-nineteen operator,
\begin{equation}
\mathcal{O}_{19}
\sim
\frac{c_{19}}{\Lambda^{15}}
\phi_{3,3}^{\dagger}
\left(udd\nu_e\right)^3
+\mathrm{h.c.},
\label{eq:dim19_operator}
\end{equation}
where $c_{19}$ is a dimensionless coefficient and $\Lambda$ is the effective suppression scale. Despite its high dimensionality, this operator can produce decays within Galactic propagation times for viable mediator masses and couplings~\cite{Gemmell:2026baryon}.

%%%%%%%%%%%%%%%%%%%%%%%%%%%%%%%%%%%%%%%%%%%%%%%%%%%%%%
\subsection{Event Generation and Antinuclei Formation}
\label{sec:coalescence}
%%%%%%%%%%%%%%%%%%%%%%%%%%%%%%%%%%%%%%%%%%%%%%%%%%%%%%

To generate events, we simulate the decay of the $\phi_{3,3}$ and $\overline{\phi}_{3,3}$ scalars using \textsc{Pythia~8.309} \cite{Bierlich:2022pfr} with the tuning parameters outlined in Ref.~\cite{DiMauro:2025vxp}. For these decays, we assume that the higher-dimensional operator produces the quarks and neutrinos promptly, with an isotropic distribution in the scalar rest frame. While this is an approximation that neglects the Lorentz structure of the off-shell fermions that have been integrated out, it provides a conservative estimate of antinuclei production by maximizing the separation of jets and thus final state antinucleons.

From these particles, we can apply a coalescence prescription to decide which antiprotons and antineutrons should bind to produce an antinucleus. In the prescription we adopt, an antinucleus forms if the constituent antinucleons are sufficiently close to each other in momentum and coordinate space; concretely, if they have a three-momentum difference smaller than $p_{\rm coal}$ and a spatial separation smaller than $d_{\rm coal}$, both measured in the rest frame of the candidate antinucleus. While other prescriptions exist, a systematic comparison found them to produce consistent spectra and multiplicities after proper calibration \cite{DiMauro:2024kml}.

The momentum and spatial cutoffs, parameterized by $p_{\rm coal}$ and $d_{\rm coal}$, are calibrated using experimental data. For antideuteron coalescence, we adopt
\begin{align}
    p_{\rm coal}^{\overline{\rm D}} = 0.219~\mathrm{GeV}, \qquad d_{\rm coal}^{\overline{\rm D}} = 3~\mathrm{fm},
\end{align}
as determined in Ref.~\cite{Gemmell:2026baryon} using ALEPH antideuteron data \cite{ALEPH:2006qoi}. For antihelium-3 coalescence, we adopt
\begin{equation}
    p_{\rm coal}^{^3\rm \overline{He}} = 0.206 \, {\rm GeV}, \qquad d_{\rm coal}^{^3 \rm \overline{He}} = 3 \, {\rm fm},
\end{equation}
as was also found in Ref.~\cite{Gemmell:2026baryon} using ALICE data \cite{ALICE:2021mfm}. The values of $p_{\rm coal}$ and $d_{\rm coal}$ used here are consistent with those obtained in Refs.~\cite{DiMauro:2024kml, DiMauro:2026oto}. 

Once the energies and momenta of the antiprotons and heavier antinuclei have been calculated in the rest frame of the decaying scalars, we apply a Lorentz boost of $m_{\rm DM}/m_\phi$ and combine the results for each scalar and antiscalar to get the spectra for a single DM annihilation in the Galactic rest frame.

%%%%%%%%%%%%%%%%%%%%%%%%%%%%%%%%%%%%%%%%%%%%%%%%%%%%%%
\section{Cosmic-ray propagation and experimental sensitivities}
\label{sec:propagation}
%%%%%%%%%%%%%%%%%%%%%%%%%%%%%%%%%%%%%%%%%%%%%%%%%%%%%%

\subsection{Galactic Propagation and Solar Modulation}

After their production in the Galactic halo, antiprotons, antideuterons, and antihelium nuclei propagate through the turbulent Galactic magnetic field before reaching the Solar System. In this section, we summarize the transport framework used to calculate their interstellar fluxes. The complete transport equation and the technical details of its solution are presented in Ref.~\cite{DiMauro:2026suep}.

We adopt the conventional axisymmetric two-zone diffusion model \cite{Strong:2007nh,Maurin:2001sj,Donato:2001ms,Genolini:2021doh,DiMauro:2021qcf}. The interstellar gas is confined to a thin disk of half-thickness $h\simeq100~\mathrm{pc}$ surrounded by a cylindrical diffusive halo. Charged particles diffuse within this magnetized volume and may also be transported away from the disk by a Galactic wind. The propagation parameters are taken from Ref.~\cite{DiMauro:2023jgg}, where they were constrained through a combined analysis of recent AMS-02 measurements of several primary and secondary cosmic-ray species.

We assume stationary transport and spatially homogeneous diffusion within the halo. Continuous energy losses and diffusive reacceleration are neglected in the calculation of the DM-induced antinucleus fluxes. This approximation is appropriate for the nuclei and energy range considered here and is discussed further in Ref.~\cite{DiMauro:2026suep}. Under these assumptions, the transport equation can be solved semi-analytically following the method introduced in Ref.~\cite{Barrau:2001ev}.

For an antimatter species $j$ produced by annihilating DM, the differential interstellar flux at the Solar position can be expressed as
\begin{equation}
\label{eq:fluxprim}
    \frac{d\Phi_j}{dK} \left(K, \mathbf{r}_\odot\right) = \frac{\beta}{4\pi} \left(\frac{\rho_\odot}{m_{\rm DM}}\right)^2 \frac{\langle\sigma v\rangle}{2} \frac{dN_j}{dK}\, \mathcal{G}_j(K).
\end{equation}
Here, $K$ denotes the kinetic energy per nucleon, $\beta=v/c$ is the velocity of the propagated particle, $m_{\rm DM}$ is the DM mass, and $\langle\sigma v\rangle$ is the velocity-averaged annihilation cross section. The quantity $dN_j/dK$ is the injection spectrum per DM annihilation taken from the results in Sec.~\ref{sec:coalescence}.

The propagation function $\mathcal{G}_j(K)$ contains the dependence on the Galactic DM distribution and the transport of species $j$. In the setup adopted here, it includes spatial diffusion, convective escape, and destructive interactions with the interstellar gas. Since the annihilation source is distributed throughout the halo, $\mathcal{G}_j$ also accounts for the convolution of the Green function of the transport equation with the squared DM density profile.

We model the Galactic DM distribution with a Navarro--Frenk--White profile \cite{1997ApJ...490..493N}, whose parameters are chosen to reproduce
\begin{equation}
\rho_\odot=0.4~\mathrm{GeV\,cm^{-3}},
\qquad
M_{200}^{\rm DM}=10^{12}\,M_\odot,
\end{equation}
where $\rho_\odot$ is the DM density at the Sun's position and $M_{200}^{\rm DM}$ is the DM halo mass enclosed in a sphere whose average density is 200 times the critical density of the universe. These values are compatible with recent determinations of the local and global Milky Way DM content \cite{2019_deSalas,deSalas:2020hbh,2025_Soding}.

The spatial diffusion coefficient is parametrized as a double-broken power law in rigidity. Following Ref.~\cite{DiMauro:2023jgg}, we place the breaks at $4~\mathrm{GV}$ and $185~\mathrm{GV}$,
and adopt rigidity slopes of
$\delta_{\rm low}=-0.80$,
$\delta_{\rm int}=0.72$,
and $\delta_{\rm high}=0.12$
below, between, and above these breaks, respectively. The diffusion coefficient is normalized at $1~\mathrm{GV}$ to
\begin{equation}
D(1~\mathrm{GV})
=
2.8\times10^{28}~\mathrm{cm^2\,s^{-1}}.
\end{equation}

For the boundary conditions of our diffusion zone, we adopt a half-thickness of 4.2 kpc~\cite{DiMauro:2023jgg}. The convective contribution is represented by a constant wind directed away from the Galactic disk,
\begin{equation}
\bm V_c
=
\mathrm{sign}(z)\,
V_c\,\hat{\bm z},
\qquad
V_c=13~\mathrm{km\,s^{-1}},
\label{eq:convective_wind}
\end{equation}
as in Ref.~\cite{Maurin:2001sj}.

Finally, the interstellar spectra are converted into top-of-atmosphere fluxes using the force-field approximation \cite{GleesonAxford:1968}. For a nucleus of charge $Z$ and mass number $A$, the interstellar and top-of-atmosphere kinetic energies per nucleon are related by
\begin{equation}
K_{\rm IS}=
K_{\rm TOA}
+
\frac{|Z|}{A}\,\phi_\odot.
\label{eq:force_field_energy}
\end{equation}
For the Fisk potential, we adopt a value of $\phi_{\odot} = 0.736 \, {\rm GV}$. The force-field treatment captures the leading effect of heliospheric modulation, although charge-sign-dependent transport and drift effects can become relevant at rigidities below a few GV \cite{John:2026ioi}.

\subsection{Antinucleus Experimental Sensitivities}
\label{sec:sensitivity}

Here, we briefly describe our assumptions about the sensitivities of AMS-02 and GAPS to $\overline{\rm D}$ and $^3\overline{\rm He}$, which are consistent with those adopted in Ref.~\cite{DiMauro:2026suep}. The projected GAPS sensitivities are based on detailed detector, reconstruction, and atmospheric simulations \cite{Aramaki:2015laa,GAPS:2020axg}. For three 35-day long-duration balloon flights, the expected antideuteron sensitivity is approximately
\begin{equation}
\phi_{\overline{\rm D}}^{\rm GAPS}
\simeq
2\times10^{-6}\,
\mathrm{(m^2\,s\,sr\,GeV/n)^{-1}}
\end{equation}
in the interval $K \in[0.1,0.3]~\mathrm{GeV/n}$. This corresponds to a $99\%$ confidence level for at least two detected events and for an expected background of about $0.014$ events. In the same energy range, the nominal $95\%$ confidence-level sensitivity to a single ${}^3\overline{\mathrm{He}}$ event is $1.3\times10^{-6}\, \mathrm{(m^2\,s\,sr\,GeV/n)^{-1}}$, although instrumental uncertainties allow for a somewhat broader range, spanning approximately $(0.06-7)\times10^{-6}$.

Early estimates of the AMS-02 antinucleus reach were obtained for a detector configuration that differs from the final flight instrument \cite{2008ICRC....4..765C}. More recent projections presented by the Collaboration indicate that, after the planned acceptance upgrade and with data accumulated through 2030, the antideuteron sensitivity could reach approximately
$8\times10^{-7}\,
\mathrm{(m^2\,s\,sr\,GeV/n)^{-1}}$
for $K \in[0.2,0.7]~\mathrm{GeV/n}$ and
$2\times10^{-7}\,
\mathrm{(m^2\,s\,sr\,GeV/n)^{-1}}$
for $K \in[2,4]~\mathrm{GeV/n}$
\cite{Oliva:JENAA:2024}.

For antihelium, the commonly used pre-launch estimate was expressed as a sensitivity to the ratio ${}^3\overline{\mathrm{He}}/\mathrm{He}\sim(1$--$2)\times10^{-9}$ \cite{Spada:2008xk}. Converting this ratio directly into an absolute flux produces a rapidly falling sensitivity with energy, as adopted in several earlier phenomenological studies \cite{Carlson:2014ssa,Cirelli:2014qia,Korsmeier:2017xzj, DeLaTorreLuque:2024htu}. Such a strong energy dependence is unlikely to be realistic, since the detector acceptance, exposure, and survival probability do not simply track the measured helium spectrum.
We therefore estimate the AMS-02 antihelium sensitivity from the predicted number of signal events per unit kinetic energy,
\begin{equation}
\phi_{\rm sens}^{\mathrm{AMS\text{-}02}}
=
\frac{dN_{\rm events}/d K}
{\mathrm{Acc} \times \mathrm{Time} \times \mathrm{Surv}},
\label{eq:ams_sens}
\end{equation}
where $\mathrm{Acc}$ is the effective acceptance, $\mathrm{Time}$ the livetime, and $\mathrm{Surv}$ the probability that the nucleus survives inelastic interactions in the detector material. Requiring at least two events gives a representative $95\%$ confidence-level sensitivity of
\begin{equation}
\phi_{{}^3\overline{\mathrm{He}}}^{\mathrm{AMS\text{-}02}}
\simeq
5\times10^{-8}\,
\mathrm{(m^2\,s\,sr\,GeV/n)^{-1}}
\end{equation}
for $K \in[2.8,10]~\mathrm{GeV/n}$.\footnote{The energy-dependent sensitivity used in this work was provided by V.~Choutko through private communication. As a consistency check, adopting a helium acceptance of $0.012~\mathrm{m^2\,sr}$~\cite{HabibyAlaoui:2016:PhD}, 12 years of exposure, and neglecting absorption in the detector yields approximately $7\times10^{-8}\, \mathrm{(m^2\,s\,sr\,GeV/n)^{-1}}$.} Outside this interval, the sensitivity deteriorates rapidly. This estimate is more conservative than the fluxes obtained by directly rescaling the pre-launch antihelium-to-helium ratio, especially above a few $\mathrm{GeV/n}$. Thus we refer to this sensitivity as conservative in Fig. \ref{fig:benchmark_fluxes}, while the rescaled ratio sensitivity is referred to as optimistic.

\begin{figure}[t]
    \centering
    \includegraphics[width=\linewidth]
    {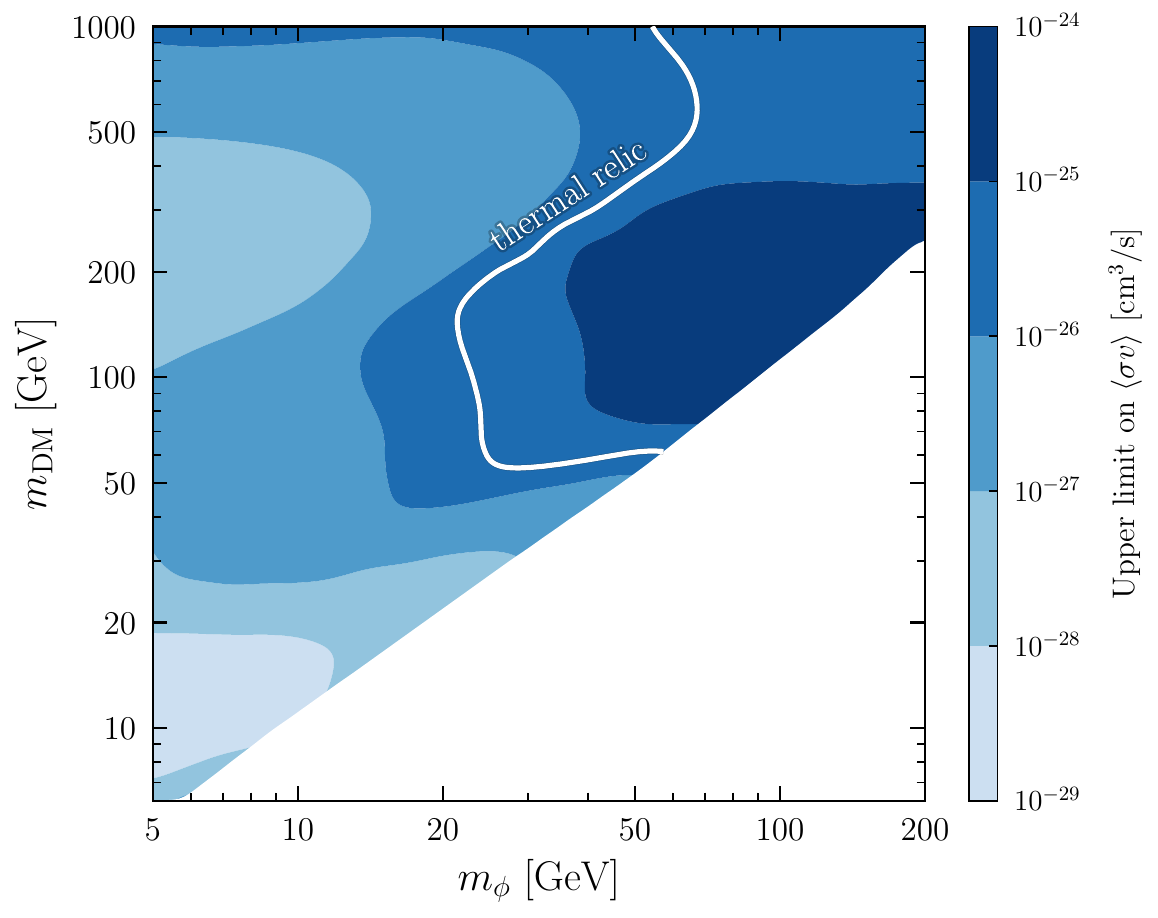}
    \caption{The 95\% confidence level upper limits on the velocity-averaged DM annihilation cross section derived from the AMS-02 antiproton  data for annihilation to $B=L=3$ scalars. The white curve illustrates where this limit is equal to the cross section of a conventional thermal relic, $\langle\sigma v\rangle = 3 \times 10^{-26}$ cm$^3$/s,
    %    
%    The white curve denotes where $\langle\sigma v\rangle_{95}^{\bar p} = \langle\sigma v\rangle_{\rm th}$
indicating that the thermal relic cross section is excluded for low $\phi$ masses, $ m_\phi\lesssim20$ GeV.}
    \label{fig:sigmav_limits}
\end{figure}

%%%%%%%%%%%%%%%%%%%%%%%%%%%%%%%%%%%%%%%%%%%%%%%%%%%%%%
\section{Results}
\label{sec:results}
%%%%%%%%%%%%%%%%%%%%%%%%%%%%%%%%%%%%%%%%%%%%%%%%%%%%%%

\subsection{Upper Limits From AMS-02 Antiproton Data}
\label{sec:results_pbar}

\begin{figure*}[t]
    \centering
    \includegraphics[width=0.485\textwidth]
    {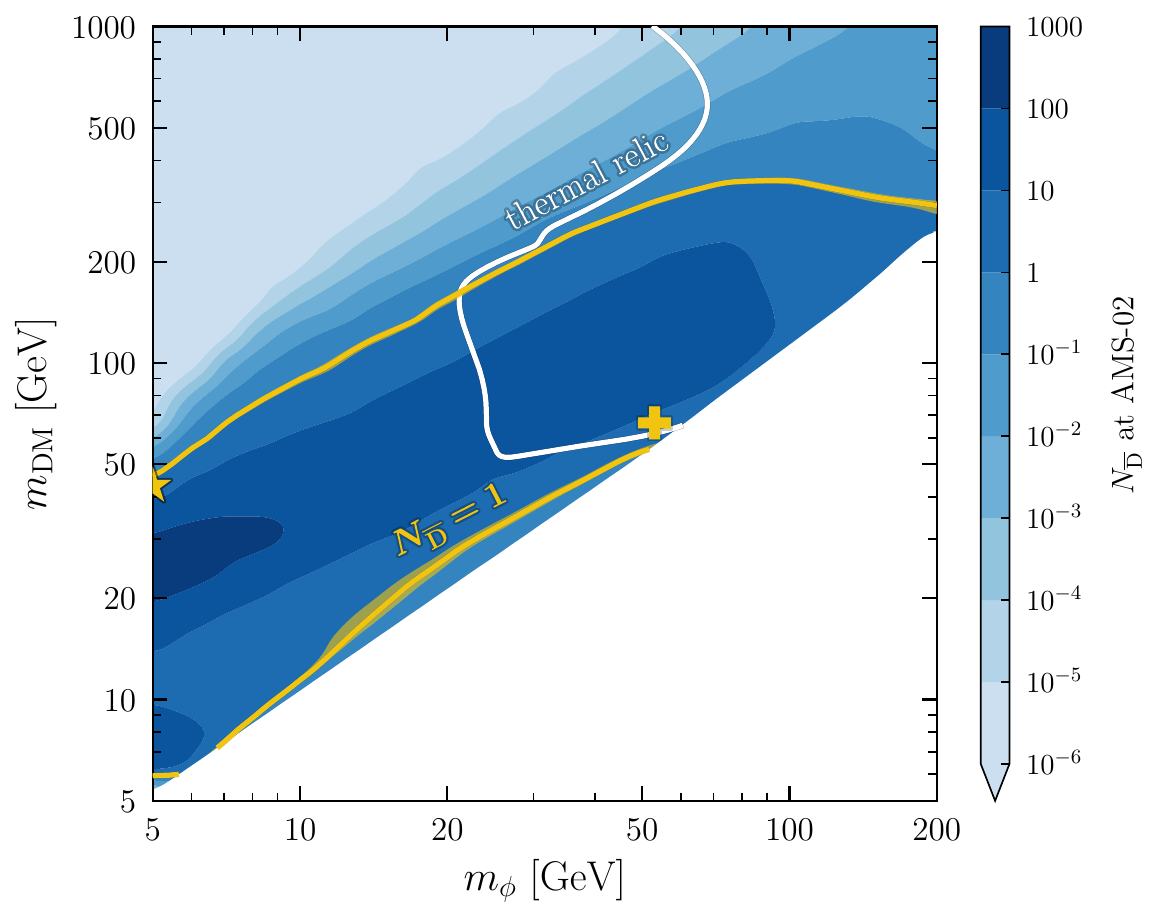}
    \hfill
    \includegraphics[width=0.485\textwidth]
    {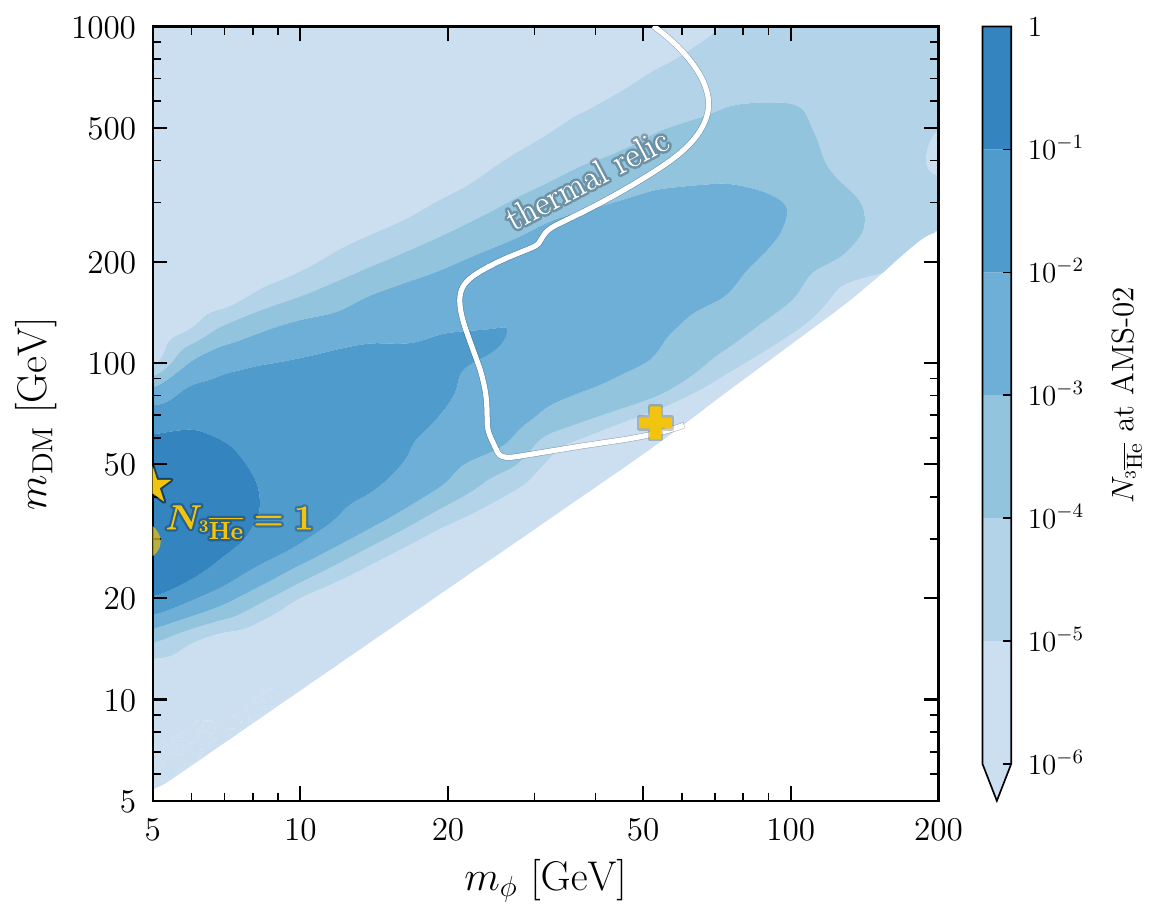}
    \caption{
    The number of antinuclei events predicted in our $B=L=3$ model, for a DM annihilation cross section that saturates the AMS-02 antiproton constraint. The left panel shows our results for $\overline{\mathrm D}$, while the right panel shows the corresponding results for ${}^3\overline{\mathrm{He}}$ using the conservative sensitivity prescription. The gold curve in each frame indicates where we find an expected number of one event at AMS-02. The white curve illustrates where the antiproton limit is equal to the cross section of a conventional thermal relic, $\langle\sigma v\rangle = 3 \times 10^{-26}$ cm$^3$/s. The \hspace{-6pt} \benchmarkstar and \hspace{-6pt} \benchmarkplus symbols denote the points in the $(m_\phi,m_{\rm DM})$ plane that are plotted in Fig.~\ref{fig:benchmark_fluxes}.
    }
    \label{fig:antinucleus_ratios_BL3_AMS02}
\end{figure*}

The production of antideuterons and antihelium nuclei is necessarily accompanied by an antiproton yield. The precise measurement of the cosmic-ray antiproton flux by AMS-02 \cite{AMS:2021nhj} therefore provides a leading constraint on the annihilation rate in the model considered here. For each point in the $(m_\phi,m_{\rm DM})$ plane, we calculate the propagated DM-induced antiproton spectrum and add it to the astrophysical secondary component taken from Ref.~\cite{DiMauro:2023jgg}.
We then determine the one-sided $95\%$ confidence-level upper limit, $\langle\sigma v\rangle_{95}^{\bar p}$, by profiling over the nuisance parameters of the antiproton model. These comprise the normalization of the secondary antiproton flux, a constant shift in its spectral index within $\pm0.1$ about a pivot rigidity of $1\,\mathrm{GV}$, and the Fisk potential describing solar modulation. The normalization and spectral-index variations account for uncertainties in the primary cosmic-ray spectra and, most importantly, in the antiproton production cross sections, which dominate the systematic uncertainty in the predicted secondary antiproton flux~\cite{diMauro:2014zea,Korsmeier:2018gcy,Maurin:2025gsz}. This procedure was also used in Ref.~\cite{DiMauro:2026suep}.

Fig.~\ref{fig:sigmav_limits} shows the resulting limits on the DM annihilation rate for the $B=L=3$ model. The white curve illustrates where this limit is equal to the cross section of a conventional thermal relic, $\langle\sigma v\rangle = 3 \times 10^{-26}$ cm$^3$/s. This limit ranges from approximately $1.5\times10^{-29}$ to $6.5\times10^{-25}\ {\rm cm^3/s}$, and is strongest for small values of $m_\phi$, as this leads to a narrow and peaked antiproton spectrum that is incompatible with that observed by AMS-02. For sufficiently large $m_{\rm DM}$ and $m_\phi$, a significant fraction of the injected antiprotons are shifted towards higher energies with a broader distribution where the constraint is less restrictive. This results in upper limits that are as weak as a few times $10^{-25}\ {\rm cm^3/s}$ in the high-mass part of the scan. In the $B=L=3$ model, it is possible to find regions with $m_\phi\gtrsim20$ GeV and $m_{\rm DM}\gtrsim50$ GeV in which the thermal relic cross section is consistent with the antiproton constraint from AMS-02.

%%%%%%%%%%%%%%%%%%%%%%%%%%%%%%%%%%%%%%%%%%%%%%%%%%%%%%
\subsection{Theoretical Predictions for
$\overline{\mathrm D}$ and ${}^3\overline{\mathrm{He}}$}
\label{sec:results_antinuclei}

In this subsection, we calculate the antinuclei fluxes that result in our model, adopting the largest values of the annihilation cross section compatible with the \mbox{AMS-02} antiproton data, as derived in Sec.~\ref{sec:results_pbar}. The resulting flux can therefore be interpreted as the maximum antinuclei signal allowed by the antiproton measurement under the assumptions adopted in this analysis. The signal scales linearly with $\langle\sigma v\rangle$, so the predictions for any smaller annihilation rate can be obtained by a direct rescaling.

Given the projected experimental sensitivity for an antinucleus species, $j$, we can calculate the predicted number of events as 
\begin{equation}
N_j(\Delta K)
=
N_{\rm sens}\Bigg(\frac{
\overline{\Phi}^{\,95}_j(\Delta K)
}{
\Phi_j^{\rm sens}(\Delta K)
}\Bigg),
\label{eq:flux_sensitivity_ratio}
\end{equation}
where $N_{\rm sens}=1$ for antihelium-3 at GAPS and 2 for antideuteron at GAPS and both antinuclei species at AMS-02, and $\Phi_j^{\rm sens}$ is the sensitivity quoted for the same kinetic energy interval and 
\begin{equation}
\overline{\Phi}^{\,95}_j(\Delta K)
=
\frac{1}{\Delta K}
\int_{\Delta K}dK\,
\frac{d\Phi_j^{95}}{dK},
\label{eq:averaged_flux}
\end{equation}
is the theoretical flux integrated and averaged over the experiment's energy coverage. For antideuterons we evaluate Eq.~\eqref{eq:flux_sensitivity_ratio} in the two projected AMS-02 search windows and sum them, whereas for ${}^3\overline{\mathrm{He}}$ we adopt the conservative AMS-02 sensitivity window.

The predicted number of antideuteron and \mbox{antihelium-3} events at AMS-02 are shown in Fig.~\ref{fig:antinucleus_ratios_BL3_AMS02}. The gold bands surrounding the $N_j=1$ contours illustrate the effect of a $\pm20\%$ uncertainty in the predicted antinuclei flux, representative of the residual uncertainty associated with the coalescence calculation.

A broad region of parameter space reaches $N_{\overline{\mathrm D}}\geq1$, indicating an observable antideuteron flux. Moreover, part of this observable region overlaps with the thermal-relic contour. The model with a thermal relic cross section can therefore produce an antideuteron signal within the reach of AMS-02 without violating the corresponding antiproton constraint.

For antihelium-3, the largest fluxes occur for light $m_\phi$ and $m_{\rm DM}$. Only in this restricted part of parameter space can the model approach the conservative AMS-02 sensitivity, while most of the region compatible with a thermal relic remains below it. Adopting some of the more optimistic sensitivity estimates used previously in the literature would improve the detection prospects, highlighting the importance of AMS-02 releasing accurate sensitivities. 

\begin{figure}[t]
    \centering
    \includegraphics[width=\columnwidth]
    {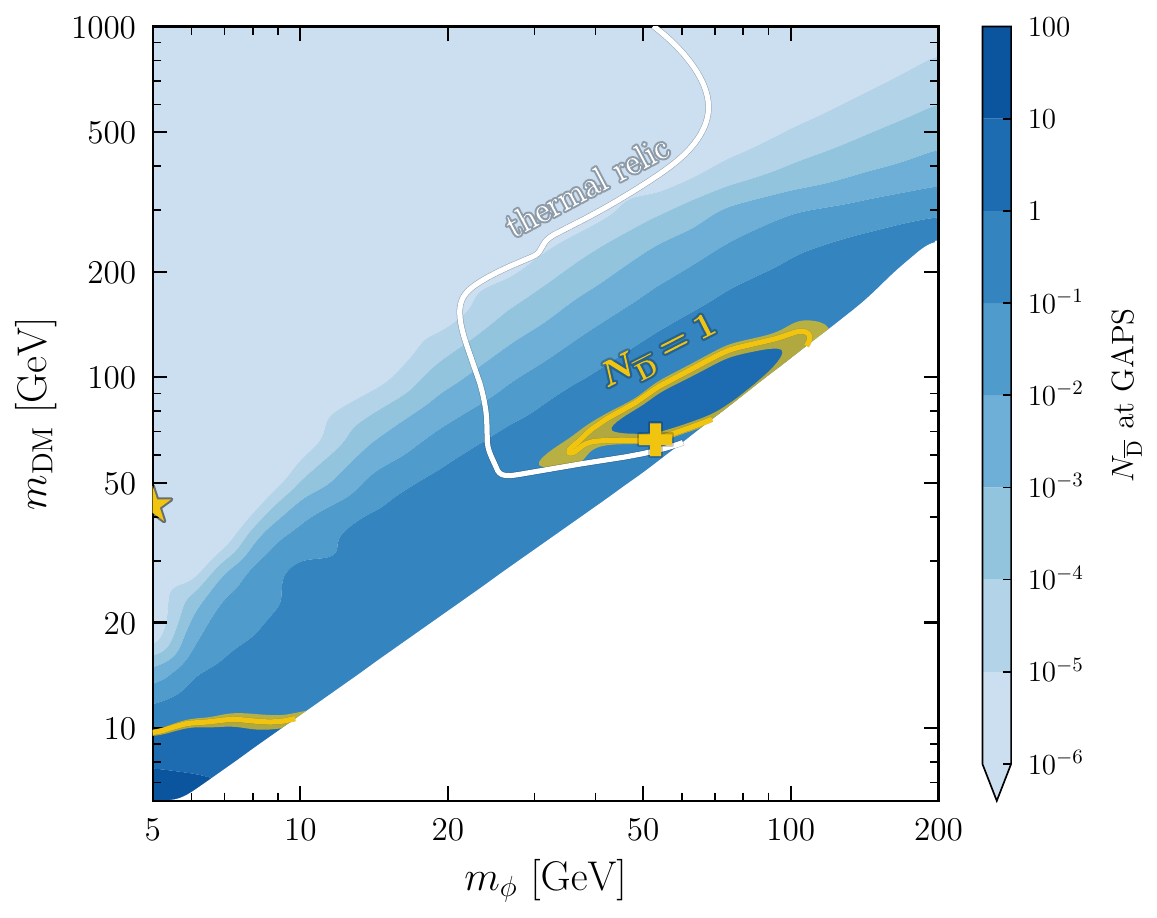}
    \caption{
    As in Fig.~\ref{fig:antinucleus_ratios_BL3_AMS02}, but for the number of antideuteron events predicted at GAPS.
    }
    \label{fig:antinucleus_ratios_BL3_GAPS}
\end{figure}

In Fig.~\ref{fig:antinucleus_ratios_BL3_GAPS}, we show the number of antideuteron events predicted at GAPS. We identify a region with DM and $\phi$ masses in the range of $50$--$100$ GeV that yields detectable signals at both GAPS and AMS-02 for an annihilation cross section that is near the thermal relic value. A smaller, separate region with $m_{\rm DM}<10$ GeV also yields a detectable antideuteron flux at GAPS. Higher antideuteron rates arise in scenarios featuring a small mass splitting between the DM and the $\phi$, so that the latter are produced with only a small boost, yielding a soft antideuteron spectrum that falls within the narrow rigidity window accessible to GAPS. We do not show corresponding contours for antihelium-3, as the predicted number of events at GAPS is orders of magnitude below unity.

\begin{figure*}[t]
    \centering
    \includegraphics[width=\textwidth]
    {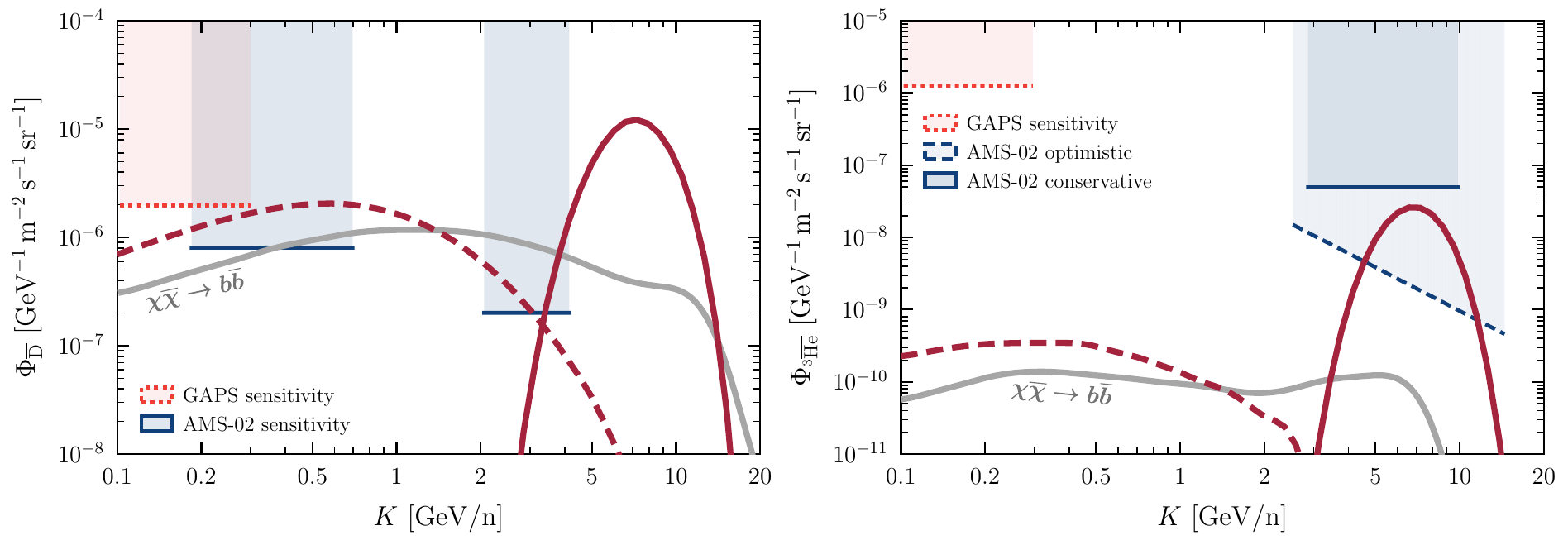}
    \caption{
    Top-of-atmosphere $\overline{\mathrm{D}}$ (left) and ${}^3\overline{\mathrm{He}}$ (right) fluxes for two representative $B=L=3$ scenarios. In each frame, the red solid line is the flux predicted for $m_{\rm DM}=43 \, {\rm GeV}$, $m_\phi = 5 \, {\rm GeV}$, and $\langle\sigma v\rangle =9.8\times10^{-28}~\mathrm{cm^3/s}$, which we labeled with a \hspace{-6pt} \benchmarkstar in Fig.~\ref{fig:antinucleus_ratios_BL3_AMS02}. The dashed line is the flux predicted for $m_{\rm DM}=66 \, {\rm GeV}$, $m_\phi = 53 \, {\rm GeV}$, and $\langle\sigma v\rangle =6.3\times10^{-26}~\mathrm{cm^3/s}$, which we labeled with a \hspace{-6pt} \benchmarkplus in the same figure. In both cases, the annihilation cross section was fixed to the 95\% confidence-level upper limit consistent with the AMS-02 antiproton data. The gray lines correspond to the fluxes predicted from DM annihilation to $b \overline{b}$ with $m_{\rm DM} = 50 ~ \mathrm{GeV}$ and $\langle\sigma v \rangle =3\times10^{-26}~\mathrm{cm^3/s}$. The projected sensitivities of AMS-02 and GAPS are included for comparison. Note that the reported sensitivity benchmarks are for \textit{two} events in each case, except for $^3\overline{\mathrm{He}}$ detection at GAPS, which is reported for a single event~\cite{2008ICRC....4..765C,Aramaki:2015laa, vonDoetinchem:2020vbj}.
    }
    \label{fig:benchmark_fluxes}
\end{figure*}

The spectral dependence of these results is illustrated in Fig.~\ref{fig:benchmark_fluxes} for two representative $B=L=3$ benchmarks. In each case, the annihilation cross section has been fixed to the upper limit derived from the antiproton spectrum measured by AMS-02. The predictions of the $B=L=3$ scalar model are compared with those of a conventional WIMP annihilating into $b\bar b$. This comparison isolates the enhancement generated by the $\phi_{3,3}$ cascade from a simple change in the overall annihilation rate.

The first benchmark has $m_{\rm DM}=43~\mathrm{GeV},~m_\phi=5~\mathrm{GeV}$, and $\langle\sigma v\rangle = 9.8\times10^{-28}~\mathrm{cm^3/s}$, chosen as a representative maximum in the antihelium-3 to antideuteron event ratio. At this mass point, we predict 3.2 antideuteron events and 0.5 antihelium-3 events at AMS-02, corresponding to a relative event ratio of 0.16, much larger than the usual $10^{-4}$ found from standard WIMP scenarios, and more in line with the $\mathcal{O}(1)$ ratio preliminarily reported by AMS-02. The resulting antideuteron spectrum is concentrated around a few $\mathrm{GeV/n}$ and reaches a maximum around $10^{-5}\, \mathrm{(m^2\,s\,sr\,GeV/n)^{-1}}$. It exceeds the conventional WIMP prediction by more than one order of magnitude and lies well above the projected AMS-02 sensitivity in the higher-energy search window. Because the spectrum is strongly suppressed below approximately $3~\mathrm{GeV/n}$, this benchmark is less favorable for the low-energy GAPS acceptance.

For the same benchmark, the ${}^3\overline{\mathrm{He}}$ spectrum peaks at a few $\mathrm{GeV/n}$ with a flux close to $10^{-7}\, \mathrm{(m^2\,s\,sr\,GeV/n)^{-1}}$. This represents an enhancement of several orders of magnitude relative to the conventional WIMP spectrum and brings the prediction nearly to the conservative \mbox{AMS-02} sensitivity while exceeding the optimistic sensitivity estimate. The present-day annihilation cross section at this benchmark is below the reference thermal value. If the same rate controlled freeze-out in a standard thermal history, the relic abundance would be too large. An additional velocity-dependent contribution to annihilation could increase the freeze-out rate while remaining suppressed today.

The second benchmark is $m_{\rm DM}=66~\mathrm{GeV},~ m_\phi=53~\mathrm{GeV}$, and $\langle\sigma v\rangle =  6.3\times10^{-26}~\mathrm{cm^3/s}$, chosen for its proximity to the thermal relic cross section. Here, the smaller scalar boost produces a broader spectrum extending to lower kinetic energies. The antideuteron flux reaches $\sim 10^{-6}\, \mathrm{(m^2\,s\,sr\,GeV/n)^{-1}}$ around $K  \simeq 0.5~\mathrm{GeV/n}$ and exceeds the projected AMS-02 sensitivity, producing an expected 9.9 antideuteron events, and an expected 1.2 events at GAPS. The corresponding ${}^3\overline{\mathrm{He}}$ flux remains below the conservative AMS-02 reach, but is still enhanced by more than an order of magnitude relative to the conventional WIMP prediction.

\subsection{Fit to the Galactic Center Gamma-Ray Excess}
\label{sec:results_gce}

The hadronic final states produced in the decay of the $B=L=3$ scalar are also accompanied by prompt gamma rays, providing an independent test of the model through {\it Fermi}-LAT observations of the Galactic Center. We therefore investigate whether the $B=L=3$ scenario can reproduce the spectrum of the Galactic Center excess (GCE). We use the GCE spectrum measured in Ref.~\cite{DiMauro:2026fnp} within a $40^\circ\times40^\circ$ region of interest around the Galactic Center, covering the energy range from $0.5$ GeV to $1$ TeV. The measured fluxes are converted into an intensity by dividing by the solid angle of the region. The prompt gamma-ray flux is calculated as
\begin{align}
    \frac{d\Phi_{\gamma}}{dE_{\gamma}} = \frac{\langle \sigma v\rangle}{8 \pi m^2_{\rm DM}} \frac{dN_{\gamma}}{dE_{\gamma}} \, J \nonumber,
\end{align}
where we define $J$ as an ROI-averaged line-of-sight integral of the DM density distribution,
\begin{align}
    J = \frac{1}{\Delta \Omega} \int_{\rm ROI} d\Omega \int_{\rm los} ds \, \rho_{\rm DM}^2,
\end{align}
and $dN_\gamma/dE_\gamma$ is the prompt photon spectrum per annihilation obtained from the same simulated events used to calculate the antimatter spectra. For the DM distribution, we adopt a generalized Navarro-Frenk-White (NFW) profile~\cite{1997ApJ...490..493N} with an inner slope of 1.3 and a scale radius of 12.7 kpc. Performing this integral over the line-of-sight and over a $40^{\circ}\times 40^{\circ}$ region of interest centered around the Galactic Center, we obtain $J =  209 \, \rho^2_{\odot} r_{\odot}$~\cite{DiMauro:2021qcf,Koechler:2025ryv}, where  $r_\odot=8.122~{\rm kpc}$ and $\rho_\odot=0.4~{\rm GeV/cm^3}$.

For each simulated $(m_{\rm DM},m_\phi)$ configuration, we profile over the annihilation cross section $\langle\sigma v\rangle$ and evaluate
\begin{equation}
\chi^2 =
\sum_i
\frac{
\left[
\Phi_i^{\rm DM}(\langle\sigma v\rangle)-\Phi_i
\right]^2
}{
\sigma_i^2
},
\label{eq:gce_chi2}
\end{equation}
where the uncertainty in each bin is selected from the lower or upper side of the systematic band depending on the sign of the residual. With one profiled normalization parameter, the fit has 10 degrees of freedom for every mass configuration.

\begin{figure}[t]
    \includegraphics[width=\columnwidth]
    {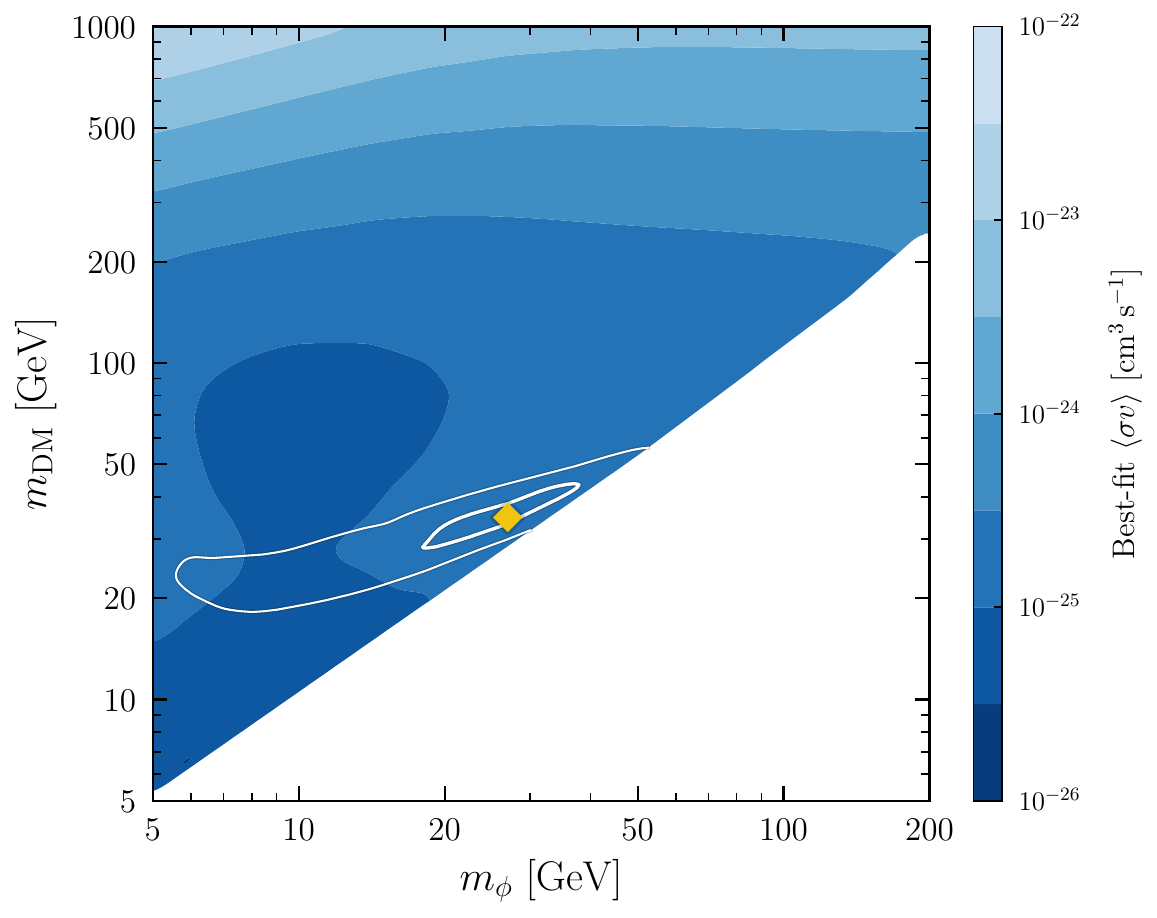}
    \caption{The best-fit value of the DM annihilation cross section from fitting the $B=L=3$ model to the Galactic Center excess in the $(m_\phi,m_{\rm DM})$ plane. The blue contours show the best-fit annihilation cross section, while the diamond marks the global best fit, located at approximately $(m_{\rm DM},m_\phi)\simeq(35,27)~{\rm GeV}$. The solid white curves show the $1\sigma$ and $2\sigma$ regions, corresponding to $\Delta\chi^2=2.30$ and $6.18$ respectively. At the global minimum, $\chi^2_{\rm min}\simeq3.46$ for 10 degrees of freedom, and for our reference DM halo profile, the preferred normalization is $\langle\sigma v\rangle\simeq1.14\times10^{-25}~{\rm cm^3/s}$. The uncolored region corresponds to $m_\phi\geq m_{\rm DM}$ and is kinematically inaccessible.
    }
    \label{fig:gce_massplane}
\end{figure}

The resulting $1 \sigma$ and $2 \sigma$ confidence contours are shown in white in Fig.~\ref{fig:gce_massplane}. The global minimum, marked with a diamond, occurs approximately at $(m_{\rm DM},m_\phi) \simeq (35~{\rm GeV},~27~{\rm GeV})$, for which $\chi^2_{\rm min}\simeq3.46$. The blue contours indicate the value of $\langle\sigma v\rangle$ preferred by the GCE at each point in the mass plane. Within the region providing the best spectral fits, the required annihilation rate is typically of order $10^{-25}~{\rm cm^3/s}$. At the global best-fit point, for the reference value, $J =  209 \, \rho^2_{\odot} r_{\odot}$, the cross section is $\langle\sigma v\rangle \simeq 1.14\times10^{-25}~{\rm cm^3/s}$.

\begin{figure}[t]
    \centering
    \includegraphics[width=\columnwidth]
    {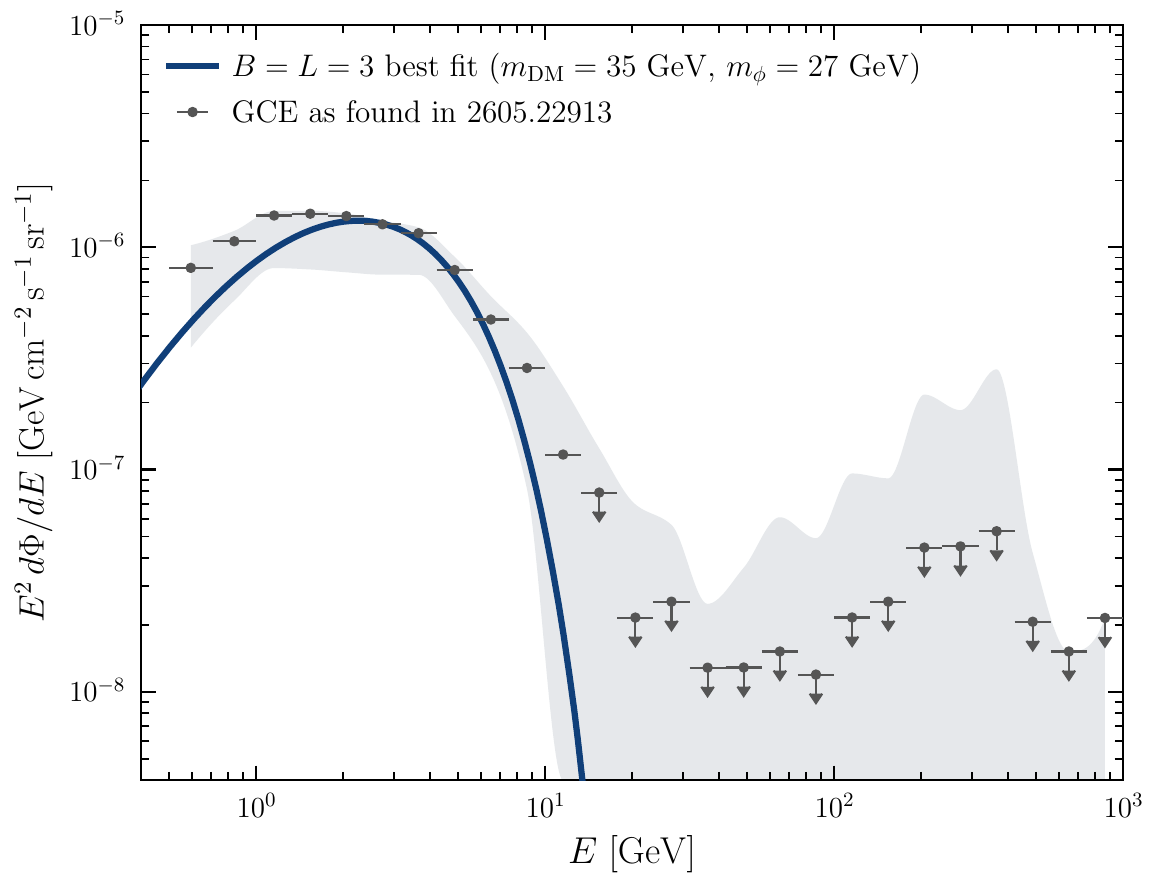}
    \caption{
    The spectral energy distribution of the Galactic Center gamma-ray excess (GCE)~\cite{DiMauro:2026fnp} compared to the prediction of our best-fit $B=L=3$ DM model. Black markers show the measured GCE fluxes, including the horizontal extent of the energy bins, while the gray region denotes the quoted systematic uncertainty band. Downward arrows show the high-energy upper limits, which are displayed for completeness but are not included in the fit. The blue solid curve is the prompt gamma-ray spectrum for the global best-fit mass configuration, $(m_{\rm DM},m_\phi)\simeq(35,27)~{\rm GeV}$. For our reference DM halo profile, the fitted normalization is $\langle\sigma v\rangle\simeq1.14\times10^{-25}~{\rm cm^3/s}$, yielding $\chi^2\simeq3.46$ for 10 degrees of freedom.
    }
    \label{fig:gce_spectrum}
\end{figure}

The best-fit spectral energy distribution is shown in Fig.~\ref{fig:gce_spectrum}. The predicted spectrum follows both the position and width of the observed GCE maximum and decreases rapidly above $10$ GeV. The agreement between the data and model demonstrates that the multi-body $B=L=3$ cascade can generate a gamma-ray signal that is consistent with the observed characteristics of the GCE.

However, the best-fit annihilation cross section required by the GCE is approximately four times the reference thermal value and exceeds the upper limit derived from the AMS-02 antiproton data. Thus, under our fiducial assumptions, this model cannot explain the GCE while remaining consistent with antiproton measurements. The larger annihilation cross section required relative to the conventional WIMP benchmark reflects the greater fraction of energy carried by protons and antiprotons, leaving less energy in unstable mesons whose decays produce gamma rays.

There is significant uncertainty in the Milky Way's DM distribution, and this could change the inferred $J$-factor by factors of order unity. Ref.~\cite{Koechler:2025ryv}, for example, found that for different halo configurations compatible with the Galactic rotation curve and the morphology of the GCE in the $40^\circ\times40^\circ$ region~\cite{DiMauro:2026fnp}, the corresponding $J$-factor can span $120 \lesssim J/\rho^2_{\odot} r_{\odot} \lesssim 422$. At fixed local density, the largest value of $J$ reduces the required annihilation cross section to approximately $5.6\times10^{-26}\,\mathrm{cm^3/s}$, closer to the thermal reference value but still above our fiducial antiproton limit. Thus, although the $B=L=3$ model provides a very good fit to the spectral shape of the GCE, a simultaneous interpretation of the GCE and cosmic-ray antimatter in terms of the same annihilating DM population is disfavored for the reference astrophysical assumptions. The comparison nevertheless illustrates the strong complementarity of the two probes. Gamma rays are particularly sensitive to the mediator-induced spectral shape and to the DM distribution in the inner Galaxy, whereas antiprotons constrain the total DM annihilation rate to hadronic channels in the local volume of the Milky Way.

%%%%%%%%%%%%%%%%%%%%%%%%%%%%%%%%%%%%%%%%%%%%%%%%%%%%%%
\section{Conclusions}
\label{sec:conclusion}
%%%%%%%%%%%%%%%%%%%%%%%%%%%%%%%%%%%%%%%%%%%%%%%%%%%%%%

Cosmic-ray antinuclei are among the most promising indirect probes of DM. In this work, we studied a recently proposed scenario in which DM annihilates into color-singlet scalar states carrying baryon and lepton number \cite{Gemmell:2026baryon}. We focused on the representative assignment $B=L=3$, for which each scalar or antiscalar decay injects three units of baryon or antibaryon number through a final state containing nine quarks or nine antiquarks, respectively. The large hadronic multiplicity and the proximity of the resulting antinucleons in momentum and position space substantially enhance the formation of antideuterons and antihelium-3 nuclei.

We calculated the antiproton, antideuteron, and antihelium-3 injection spectra using an event-by-event Monte Carlo treatment of hadronization and coalescence, with coalescence parameters calibrated to collider data. We then propagated these spectra through the Galaxy using a transport model constrained by recent measurements of primary and secondary cosmic-ray species. At each sampled point in the $(m_\phi,m_{\rm DM})$ plane, we derived a $95\%$ confidence-level upper limit on the annihilation cross section using the seven-year AMS-02 antiproton data.

Setting the annihilation cross section to its AMS-02 antiproton upper limit, we find antideuteron fluxes within the projected reach of AMS-02 and GAPS across substantial regions of parameter space. Detectable antideuteron signals are possible for a DM annihilation cross section equal to that of a thermal relic.

Although the $B=L=3$ decay can enhance the antihelium-3 flux by several orders of magnitude relative to conventional WIMP annihilation, its detection prospects at AMS-02 remain marginal under our conservative sensitivity assumptions. The largest fluxes arise for light mediators and moderately light DM. In this region, the predicted ${}^3\overline{\mathrm{He}}$ flux can approach the projected AMS-02 sensitivity. For example, at our benchmark point $(m_{\rm DM},m_\phi)=(43,5)~\mathrm{GeV}$, we predict 0.5 antihelium-3 events and 3.2 antideuteron events at AMS-02. Their ratio of approximately 0.16 is substantially larger than in conventional WIMP scenarios and closer to the relative abundance suggested by the reported AMS-02 candidate events.

We also investigated the prompt gamma-ray emission associated with the $B=L=3$ cascade and found that it provides a good fit to the spectral shape of the Galactic Center gamma-ray excess, with a global best fit near $m_{\rm DM} \approx 35~\mathrm{GeV}$ and $m_\phi \approx 27~\mathrm{GeV}$. However, for our reference halo profile, the required annihilation cross section exceeds the AMS-02 antiproton upper limit at these masses. Thus, despite the good spectral agreement, a simultaneous interpretation of the gamma-ray excess and cosmic-ray antimatter signals in terms of the same annihilating DM population is disfavored under our astrophysical assumptions.

The main uncertainties in the results presented here arise from our models of antinucleus coalescence, Galactic transport, solar modulation, and the antiproton background. Although our coalescence parameters are calibrated to collider measurements, future measurements of antideuteron and antihelium production in different collider environments will be essential for testing the universality of the formation model. Improved measurements of antiproton production cross sections and tighter constraints on cosmic-ray transport parameters will likewise reduce the uncertainties in the inferred upper limits on the annihilation rate.

To conclude, the direct injection of multiple units of antibaryon number provides an efficient mechanism for overcoming the strong suppression of antinucleus production characteristic of conventional WIMP annihilation. The $B=L=3$ cascade can yield observable antideuteron fluxes and, in favorable regions of parameter space, potentially detectable antihelium-3 fluxes, with antihelium-to-antideuteron event ratios reaching $\mathcal{O}(10^{-1})$, all while remaining consistent with the AMS-02 antiproton data. Clarifying the nature of the reported AMS-02 candidate events and establishing the instrument's in-flight sensitivity to antihelium-3 will be crucial for further assessing this interpretation. Antideuteron searches with GAPS will provide a sensitive and complementary test of dark sectors carrying baryon number.

\begin{acknowledgements}
We thank Seth Koren for helpful conversations. The work of CG, DH and FV has been supported by the Office of the Vice Chancellor for Research at the University of Wisconsin-Madison, with funding from the Wisconsin Alumni Research Foundation. MDM acknowledges support from the research grant {\sl TAsP (Theoretical Astroparticle Physics)} funded by Istituto Nazionale di Fisica Nucleare (INFN).
\end{acknowledgements}

\appendix

\begin{figure}[!b]
    \centering
    \includegraphics[width=\linewidth]
    {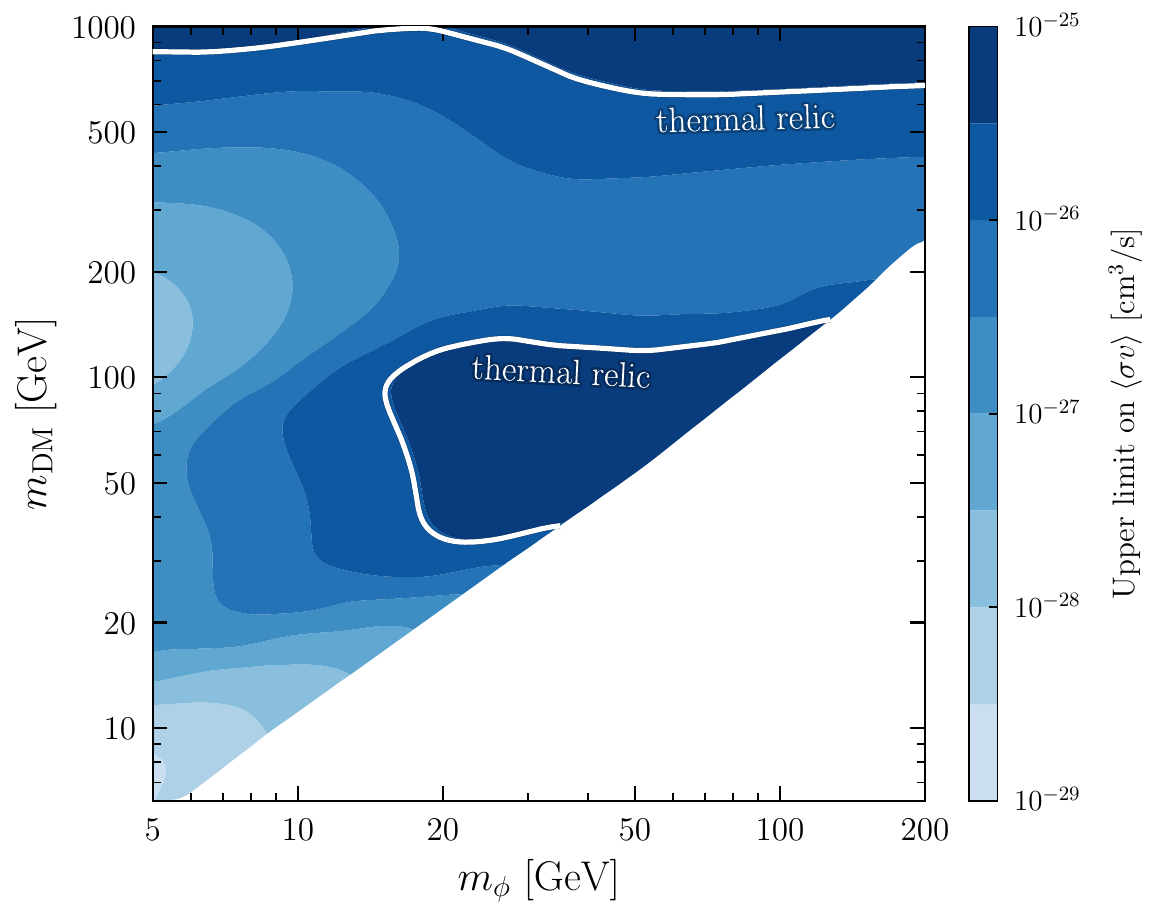}    
    \caption{The 95\% confidence level upper limits on the velocity-averaged DM annihilation cross section derived from the AMS-02 antiproton  data for annihilation to $B=L=1$ scalars. The white curve illustrates where this limit is equal to the cross section of a conventional thermal relic, $\langle\sigma v\rangle = 3 \times 10^{-26}$ cm$^3$/s.}
    \label{app:sigmav_limits}
\end{figure}

\section{Results for $B=L=1$}
\label{app:other_scalars}

While the $B=L=1$ scenario does not produce an enhancement in the antihelium-3 flux compared to the WIMP scenario, it can still lead to an enhanced antideuteron flux \cite{Gemmell:2026baryon}. In this appendix, we show the predicted number of $\overline{\rm D}$ events at AMS-02 for this alternate model, calculated using the same procedure as for the $B=L=3$ case.

In Fig. \ref{app:sigmav_limits}, we present the maximum allowed annihilation cross section consistent with AMS-02 antiproton measurements. The limits span approximately $2.4\times10^{-29} \lesssim \langle\sigma v\rangle_{95}^{\bar p} \lesssim 7.2\times10^{-26} \ {\rm cm^3/s}$ over the mass range investigated. The strongest constraints are found in the low-$m_\phi$ region, for a range of DM masses, where the injected antiprotons populate the energy interval most accurately measured by AMS-02. At larger masses, the bound depends non-monotonically on $m_\phi$, reflecting the variation of both the scalar boost and the antiproton spectrum generated in its decay.

\begin{figure}[!t]
    \centering
    \includegraphics[width=0.485\textwidth]
    {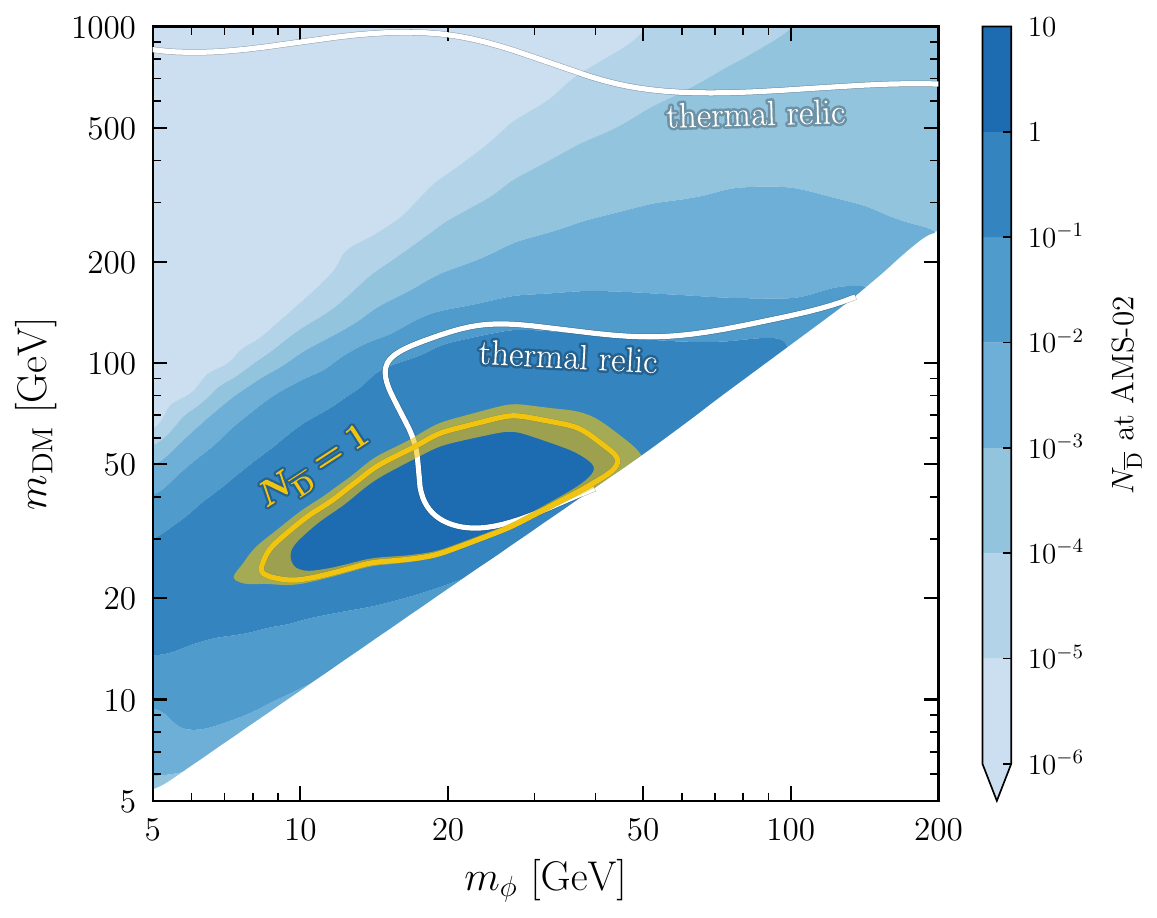}
    \caption{
    As in Fig.~\ref{fig:antinucleus_ratios_BL3_AMS02}, but for the number of $\overline{\rm D}$ events predicted at AMS-02 in the $B=L=1$ scenario.}
    \label{fig:antinucleus_ratios_BL1}
\end{figure}

Fig.~\ref{fig:antinucleus_ratios_BL1} shows the predicted number of antideuteron events at AMS-02. For 20 GeV $\lesssim m_{\rm DM} \lesssim$ 50 GeV and $m_\phi \sim 20$ GeV, the $B=L=1$ scenario can yield approximately two events in regions where a thermal annihilation cross section remains allowed. 

\bibliography{main}

\end{document}